\documentclass[aps,prd,showpacs,amssymb,nofootinbib,twocolumn]{revtex4}

\usepackage{amsmath}
\usepackage{amssymb}
\usepackage{latexsym}
\usepackage{amsfonts}
\usepackage{epsfig}
\usepackage{psfrag}
\usepackage{graphicx}

\usepackage{color}

\definecolor{black}{rgb}{0,0,0}
\definecolor{blue}{rgb}{0,0,1}
\definecolor{green}{rgb}{0,1,0}
\definecolor{red}{rgb}{1,0,0}
\definecolor{brown}{rgb}{0.4,0.2,0}
\definecolor{darkgreen}{rgb}{0,0.7,0}

\usepackage{bm}
\usepackage{amsmath}
\usepackage{amssymb}
\usepackage{latexsym}
\usepackage{amsfonts}
\usepackage{epsfig}
\usepackage{psfrag}
\usepackage{graphicx}

\newcommand{\ket}[1]{\left|#1\right>}
\newcommand{\bra}[1]{\left<#1\right|}

\newcommand{\scalar}[2]{\left\{#1|#2\right\}}

\newcommand{\inner}[2]{\left(#1|#2\right)}

\newcommand{\nn}{\nonumber\\}

\newcommand{\f}[1]{\mbox{\boldmath$#1$}}
\newcommand{\fk}[1]{\mbox{\boldmath$\scriptstyle#1$}}

\newcommand{\bea}{\begin{eqnarray}}
\newcommand{\ea}{\end{eqnarray}}
\newcommand{\eea}{\end{eqnarray}}
\newcommand{\ord}{{\cal O}}

\begin{document}

%%%%%%%%%%%%%%%%%%%%%%%%%%%%%%%%%%%%%%%%%%%%%%%%%%%%%%%%%%%%%%%%%%%%%%%%%%%%%%
\title{
On the partner particles for moving mirror radiation and black hole evaporation}
%%%%%%%%%%%%%%%%%%%%%%%%%%%%%%%%%%%%%%%%%%%%%%%%%%%%%%%%%%%%%%%%%%%%%%%%%%%%%%

\author{M.~Hotta$^1$}
\email[e-mail:\,]{hotta@tuhep.phys.tohoku.ac.jp}
\author{ R.~Sch\"utzhold$^2$}
\email[e-mail:\,]{ralf.schuetzhold@uni-due.de}
\author{W.~G.~Unruh$^3$}
\email[e-mail:\,]{unruh@physics.ubc.ca}

\affiliation{
$^1$Graduate School of Science, Tohoku University, Aobaku, Aramaki, 
Aza 6-3, Sendai, 980-8678, Japan
\\
$^2$Fakult\"at f\"ur Physik, Universit\"at Duisburg-Essen, 
Lotharstrasse 1, 47057 Duisburg, Germany
\\
$^3$CIAR Cosmology and Gravity Program, 
Dept.\ of Physics, University of B.C., Vancouver, Canada V6T 1Z1}

\date{\today}

\begin{abstract}
The partner mode with respect to a vacuum state for a given mode (like
that corresponding to one of the thermal particles emitted by a black hole) 
is defined and calculated. 
The partner modes are explicitly calculated for a number of cases, 
in particular for the modes corresponding to a particle detector being 
excited by turn-on/turn-off transients, or with the thermal particles 
emitted by the accelerated mirror model for black hole evaporation. 
One of the key results is that the partner mode in general is just a vacuum 
fluctuation, and one can have the partner mode be located in a region where 
the state cannot be distinguished from the vacuum state by any series of 
local measurements, including the energy density. 
I.e., "information" (the correlations with the thermal emissions) need not be 
associated with any energy transport. 
The idea that black holes emit huge amounts of energy in their last stages 
because of all the information which must be emitted under the assumption 
of black-hole unitarity is found not necessarily to be the case.
\end{abstract}


\maketitle

%%%%%%%%%%%%%%%%%%%%%%%%%%%%%%%%%%%%%%%%%%%%%%%%%%%%%%%%%%%%%%%%%%%%%%%%%%%%%%
\section{Introduction}
%%%%%%%%%%%%%%%%%%%%%%%%%%%%%%%%%%%%%%%%%%%%%%%%%%%%%%%%%%%%%%%%%%%%%%%%%%%%%%

Quantum mechanics around a black hole has been one of the most exciting and
puzzling aspects of theoretical physics in the past half century 
\cite{hawking}. 
One of the issues has been that of "information" and how information is carried.
Black-hole unitarity \cite{unitarity} is the belief that in the space-time 
outside the black hole, the evolution of a quantum field from a time before 
the black hole formed to after the black hole completely evaporated must be 
unitary -- initial states of the quantum field map uniquely to final states 
of the quantum field. 
If one believes in "black hole unitarity" (and in this paper we are agnostic 
about that belief) then there must exist correlations between the early 
Hawking evaporation emission from the black hole, and late time emission. 
For any particle emitted early on, some correlation between this early 
mission and the field later on must exist. 
Given a mode which carries away thermal particles in the early stages, 
there must be "partner modes" which occur later which are correlated with 
these early modes in order  that "unitarity" be preserved. 

The characterization of these partners thus becomes important. 
In section~\ref{Definition-Partner} we define the partner mode, 
uniquely in some cases of pure two-mode squeezing. 
In section~\ref{Partners and detectors} we show that any particle detection 
measurement of the field also has a partner, even in the case where the 
detector is stationary but is switched on and off. 
In section~\ref{Moving Mirror Radiation} we look at the partner in the case 
of the accelerated mirror model of black hole thermal emission. 

One of the surprising results is that the partner need not be located near the
original mode, but can be located in distant regions of the space-time. 
While recognized in the correlations between the field inside and outside the 
black hole in the Hawking evaporation process \cite{SU}, this is a general 
feature of the partner modes.
Our results are somewhat related to to recent observations that the long range
entanglement in the vacuum can be used to entangle other systems even in
spatially separated regions \cite{reznik} and to energy teleportation studies 
\cite{hotta}.

First, let us specify how the partner mode can be defined.
To this end, we demand the two conditions: 

{\bf A}
The reduced density matrix of the Hawking plus partner modes obtained by 
integrating out all other degrees of freedom should be a pure state.
Since the total state (the initial vacuum) is a pure state, this is equivalent 
to vanishing entanglement between the Hawking plus partner mode 
on the one hand and the rest of the system on the other hand. 

However, this requirement alone does not define the partner mode uniquely 
(see below).
For example, one could envisage a single-mode squeezing operation
and phase rotation acting on the partner mode, which does not change 
the purity of the combined state (Hawking plus partner). 
Specifying the partner mode uniquely requires a second condition. 
There are several reasonable options, here we list some possibilities:  

{\bf B1}
The quantum state after absorbing (annihilating) one partner particle 
should be (up to normalization due to possibly different probabilities)
the same state as after creating one Hawking particle.
This corresponds to the intuitive picture that the Hawking and partner 
particles always come in pairs. 

Alternatively, one could implement the idea that  Hawking and partner 
particles always come in pairs by imposing the requirement the 
other way around: 

{\bf B2}
The quantum state after absorbing one Hawking particle should be 
(again up to normalization)
the same state as after creating one partner particle.

As we shall see below, condition {\bf B1} can always be satisfied -- 
unless the Hawking mode contains single-mode squeezing only and 
thus there would be no need for a partner particle at all -- 
whereas the requirement {\bf B2} can only be fulfilled if the 
single-mode squeezing of the Hawking mode is small enough. 
As another option, we could demand that the probabilities for detecting 
Hawking and partner particles should be the same -- treating these 
two modes on a symmetric footing. 
As it turns out, in the scenarios we are interested in below 
(pure two-mode squeezing), all these requirements yield the same 
answer for the partner particle. 

%%%%%%%%%%%%%%%%%%%%%%%%%%%%%%%%%%%%%%%%%%%%%%%%%%%%%%%%%%%%%%%%%%%%%%%%%%%%%%
\section{Definition of Partner Particle}\label{Definition-Partner}
%%%%%%%%%%%%%%%%%%%%%%%%%%%%%%%%%%%%%%%%%%%%%%%%%%%%%%%%%%%%%%%%%%%%%%%%%%%%%%

Now let us show how to satisfy these requirements. 
As a most general ansatz, we decompose the Hawking mode 
\bea
\hat a_H=\int dk\left(\alpha_k^*\hat a_k+\beta_k\hat a_k^\dagger\right)
\,,
\ea
into creation and annihilation operators $\hat a_k^\dagger$ and $\hat a_k$ 
\bea
\left[\hat a_k,\hat a_{k'}\right]=
\left[\hat a_k^\dagger,\hat a_{k'}^\dagger\right]=0
\,,\;
\left[\hat a_k,\hat a_{k'}^\dagger\right]=
\delta(k,k')
\,,
\ea
defined with respect to the initial vacuum state 
\bea
\forall_k\:\hat a_k\ket{0}=0 
\,,
\ea
where $k$ denotes some quantum number.

For convenience, let us introduce the usual complex scalar product of 
two functions or vectors $\f{\chi}$ and $\f{\zeta}$ via 
\bea
\scalar{\f{\chi}}{\f{\zeta}}=\int dk\,\chi_k^* \zeta_k
\,.
\ea
Accordingly, we define the projection of the initial annihilation operators 
$\hat a_k$ onto one mode $\f{\chi}$ via 
\bea
\hat a_{\fk{\chi}}=
\scalar{\f{\chi}}{\hat{\f{a}}}=\int dk\, \chi_k^* \hat a_k
\,,
\ea
which gives the commutation relations 
\bea
\label{commutators}
\left[\hat a_{\fk{\chi}},\hat a_{\fk{\zeta}}\right]=
\left[\hat a_{\fk{\chi}}^\dagger,\hat a_{\fk{\zeta}}^\dagger\right]=0
\,,\;
\left[\hat a_{\fk{\chi}},\hat a_{\fk{\zeta}}^\dagger\right]=
\scalar{\f{\chi}}{\f{\zeta}}
\,.
\ea
In this notation, the Hawking mode is given by 
\bea
\hat a_H
=
\scalar{\f{\alpha}}{\hat{\f{a}}}+
\left(\scalar{\f{\beta}}{\hat{\f{a}}}\right)^\dagger
=
\scalar{\f{\alpha}}{\hat{\f{a}}}+
\scalar{\hat{\f{a}}}{\f{\beta}}
\,.
\ea
Now let us introduce an orthonormal basis $\f{n}_\|$ and $\f{n}_\perp$ 
in the subspace spanned by the two vectors $\f{\alpha}$ and $\f{\beta}$
\bea
\f{\alpha}=\alpha\f{n}_\|
\,,\quad
\f{\beta}=\beta_\|\f{n}_\|+\beta_\perp\f{n}_\perp 
%\,,\quad
%\f{\alpha}=\alpha_\|  \f{n}_\| + \alpha_\perp  \f{n}_\perp
\,,
\ea
where $|\f{n}_\||^2=\scalar{\f{n}_\|}{\f{n}_\|}=1=|\f{n}_\perp|^2$ and 
$\f{n}_\|\perp\f{n}_\perp$, i.e.,  $\scalar{\f{n}_\|}{\f{n}_\perp}=0$.
If $\beta_\perp=0$, we would have pure single-mode squeezing and 
the Hawking mode itself would be in a pure state, i.e., there would be 
no need for a partner particle. 
In the general case $\beta_\perp\neq0$, we can restrict ourselves 
to the two modes 
\bea
\hat a_\|=\scalar{\f{n}_\|}{\hat{\f{a}}}
\,,\;
\hat a_\perp=\scalar{\f{n}_\perp}{\hat{\f{a}}}
\,,
\ea
which satisfy the usual commutation relations due to 
Eq.~(\ref{commutators}).
These operators annihilate the initial vacuum 
\bea
\hat a_\|\ket{0}=\hat a_\perp\ket{0}=0
\,,
\ea
and thus the reduced density matrix of these two modes is a pure state. 
Now the idea is that everything involving the Hawking mode $\hat a_H$ 
and its partner mode $\hat a_P$ will occur in the two-mode space
spanned by $\hat a_\|$ and $\hat a_\perp$ and their adjoints 
$\hat a_\|^\dagger$ and $\hat a_\perp^\dagger$.
As we show the the Appendix, this is actually the only way to 
satisfy requirement {\bf A}. 
In terms of these operators, the Hawking mode is given by 
\bea
\hat a_H=\alpha^*\hat a_\|
%+\alpha_\perp^*\hat a_\perp
+\beta_\|\hat a_\|^\dagger+\beta_\perp\hat a_\perp^\dagger
\,.
\ea
From $[\hat a_H,\hat a_H^\dagger]=1$ follows 
$|\alpha|^2-|\beta_\||^2-|\beta_\perp|^2=1$. 
Note that one can make the Bogoliubov coefficients 
$\alpha$, $\beta_\|$, and $\beta_\perp$ real by 
absorbing their phases into  $\hat a_\|$, $\hat a_\perp$, 
and $\hat a_H$. 

Following our strategy, we can make the following general ansatz 
for the partner particle 
\bea
\label{partner-ansatz}
\hat a_P
=
\gamma_\|^*\hat a_\|+
\gamma_\perp^*\hat a_\perp+
\delta_\|\hat a_\|^\dagger+
\delta_\perp\hat a_\perp^\dagger
\,.
\ea
In this way, requirement {\bf A} is automatically satisfied. 
Since $\hat a_P$ should obey the usual commutation relation 
$[\hat a_P,\hat a_P^\dagger]=1$,  the  above Bogoliubov coefficients 
should satisfy $|\gamma_\||^2+|\gamma_\perp|^2-|\delta_\||^2-|\delta_\perp|^2=1$.
Furthermore, since we want the two modes $\hat a_H$ and $\hat a_P$
to be independent, i.e., 
$[\hat a_P,\hat a_H^\dagger]=0=[\hat a_H,\hat a_P^\dagger]$ as well as 
$[\hat a_P,\hat a_H]=0$, we get the conditions 
$\gamma_\|^*\alpha=\beta_\|^*\delta_\|+\beta_\perp^*\delta_\perp$ 
and 
$\gamma_\|^*\beta_\|+\gamma_\perp^*\beta_\perp=\alpha^*\delta_\|$. 

As mentioned above, these three equations do not specify the four 
Bogoliubov coefficients for $\hat a_P$ uniquely. 
We could still apply a single-mode squeezing/phase transformation 
$\hat a_P\to e^{i\varphi}\cosh\zeta\,\hat a_P+
e^{i\vartheta}\sinh\zeta\,\hat a_P^\dagger$ 
within the $\hat a_P$-mode for arbitrary (real) values of 
$\varphi$, $\zeta$, and $\varphi$ 
without violating any of the conditions above. 
In order to fix this remaining degree of freedom, a second requirement  
is necessary -- here, we discuss {\bf B1} and {\bf B2}.  

Option {\bf B1} corresponds to choosing 
$\hat a_P\ket{0}\propto\hat a_H^\dagger\ket{0}$, i.e., 
$\f{\delta}\|\f{\alpha}$, 
which means $\delta_\perp=0$ and $\delta_\|=\delta$. 
This then gives $\gamma_\|^*= \beta_\|^*\delta/\alpha$ and 
$\gamma_\perp^*=(\beta_\perp^{-1}+\beta_\perp^*)\delta/\alpha$
such that the remaining Bogoliubov coefficient $\delta$ can be 
determined by the unitarity condition 
$|\f{\gamma}|^2-|\f{\delta}|^2=1$ 
up to a global phase. 
Writing this {\bf B1}-condition $\hat a_P\ket{0}\propto\hat a_H^\dagger\ket{0}$
in the form $\hat a_P\ket{0}=\eta\hat a_H^\dagger\ket{0}$ with some constant 
$\eta$, we have 
\bea
\left(\hat a_P-\eta\hat a_H^\dagger\right)\ket{0}=0
\,.
\ea
Thus, this linear combination $\hat a_P-\eta\hat a_H^\dagger$
is composed of initial annihilation operators only. 

The other option {\bf B2} corresponds to 
$\hat a_H\ket{0}\propto\hat a_P^\dagger\ket{0}$, i.e., 
$\f{\gamma}\|\f{\beta}$, which allows us to determine the 
Bogoliubov coefficients in a completely analogous manner.
Note, however, that there is an important difference:
As shown above, condition {\bf B1} can always be fulfilled unless 
\mbox{$\beta_\perp=0$}, in which case we would have pure 
single-mode squeezing within the $\hat a_H$-mode and there 
would be no need for a partner mode. 
In contrast, requirement {\bf B2} cannot be satisfied 
if the amount of single-mode squeezing becomes too large.  

In case of vanishing  single-mode squeezing $\beta_\|=0$, 
both requirements give the same partner mode
\bea
\hat a_P=\alpha^*\hat a_\perp+
\beta\hat a_\|^\dagger
\,,
\ea
where $\alpha$ and $\beta$ can be made real because their phases can 
be absorbed into the definition of $\hat a_\|$ and $\hat a_\perp$.
In this case, the initial vacuum state restricted to the two modes 
$\hat a_H$ and $\hat a_P$ is a pure two-mode squeezed state 
\bea
\label{squeezing}
\ket{0}=
\exp\left\{ \xi\, \hat a_H^\dagger \hat a_P^\dagger - {\rm h.c.} \right\}
\ket{0}_{HP}
\,,
\ea
with respect to the zero-particle state $\ket{0}_{HP}$ which is annihilated 
by $\hat a_H$ and $\hat a_P$ 
\bea
\hat a_H\ket{0}_{HP}=\hat a_P\ket{0}_{HP}=0
\,,
\ea
where the squeezing parameter $\xi$ satisfies $\alpha=\cosh\xi$ and 
$\beta=\sinh\xi$.
As a result, the initial vacuum $\ket{0}$ can be viewed as a state 
containing pairs of particles in the modes $\hat a_H$ and $\hat a_P$.
After tracing out (averaging over) the partner mode, this squeezed 
state~(\ref{squeezing}) yields a thermal type density matrix for the 
Hawking mode 
\bea
\hat\rho_H=
\frac{1}{Z}\,\exp\left\{- \frac{\hat a_H^\dagger\hat a_H}{{\cal T}_H}\right\}
\,,
\ea
with the normalization %(partition function) 
$Z$ ensuring ${\rm Tr}\{\hat\rho_H\}=1$
and 
\bea
{\cal T}_H=\frac{1}{2\ln(\cosh\xi)}
\,,
\ea
which can be regarded as a dimensionless Hawking temperature.  
Due to the symmetric nature of the squeezed state~(\ref{squeezing}),
the same applies to the reduced state of the partner particles 
(after tracing out the Hawking mode).

Note that the mapping from the Hawking mode $\hat a_H$ to its partner mode 
$\hat a_P$ is not linear in general -- if 
$\hat a_H$ has the partner mode $\hat a_P$ 
and 
$\hat a_H'$ has the partner mode $\hat a_P'$,
then the partner mode for $\mu\hat a_H+\nu\hat a_H'$, for example, 
is almost never $\mu\hat a_P+\nu\hat a_P'$, even if we have 
no single-mode squeezing in both cases.

%%%%%%%%%%%%%%%%%%%%%%%%%%%%%%%%%%%%%%%%%%%%%%%%%%%%%%%%%%%%%%%%%%%%%%%%%%%%%%%%
\section{Partners and detectors}\label{Partners and detectors}
%%%%%%%%%%%%%%%%%%%%%%%%%%%%%%%%%%%%%%%%%%%%%%%%%%%%%%%%%%%%%%%%%%%%%%%%%%%%%%%%

The idea of a partner particle has a broader applicability than just Hawking 
or acceleration (Unruh) radiation. 
Consider a model particle detector as suggested by Unruh \cite{unruh} and 
developed by De~Witt \cite{dewitt}. 
The detector is taken as occupying a single point in space-time with an 
internal degree of freedom, often taken to be a spin degree, 
but could equally and more simply be taken to be a harmonic oscillator 
degree of freedom. 
The energy difference (in the rest frame of the detector) between the ground 
state and the first excited state is $E$.
This is coupled to the quantum field of interest. 
This detector responds to specific degrees of freedom of the field, changing 
its state from ground to excited state, which is regarded as a detection. 
(I.e., if the detector is discovered at some time to be in its excited
state, it must have absorbed energy and a particle from the field.) 

The interaction Lagrangian is given by ($\hbar=c=1$) 
\bea
L_{\rm int}
=
\epsilon(\tau)\,q(\tau)\,\partial_\tau\Phi[t(\tau),x(\tau)]
\,,
\eea
where $\epsilon(\tau)$ is the possibly time dependent coupling,
$t(\tau),x(\tau)$ is the trajectory of the detector in terms of the proper
time along the path $\tau$.
After quantization, the internal degree of freedom of the detector 
corresponds to the operator
\bea
\hat q(\tau)= \hat c e^{-iE\tau} +\hat c^\dagger e^{iE\tau}
\,,
\eea
where $\hat c$ is the annihilation operator taking the detector from the 
first excited state of energy $E$ to the ground state of zero energy. 
Note that if the detector is a harmonic oscillator, then $\hat c$ could just 
be $\sqrt{2E}$ times the usual oscillator annihilation operator. 
The normalization of $\hat c$ is not important because it will cancel out in 
the following anyway.

One can define a field  operator associated with the detector by
\bea
\hat a_D = {\cal N} 
\int d\tau\,\epsilon(\tau)\,e^{iE\tau}\, 
\partial_\tau\hat\Phi[t(\tau), x(\tau)]
\,, 
\eea
where $\cal N$ is chosen so as to make
\bea
\left[\hat a_D,\hat a_D^\dagger\right]=1
\,.
\eea
Furthermore, one can define a mode function associated with this operator by 
\bea
\phi_D(t,x)= \left[\hat\Phi(t,x),\hat a_D^\dagger\right]
\,.
\eea
Assuming that the field is (initially) in a vacuum state, 
the excitation probability of the detector will be given by
\bea
{\cal P}= {\bra{0}\hat a_D^\dagger\hat a_D\ket{0}\over{\cal N}^2}
\,.
\eea
Hence $\hat a_D$ corresponds to the Hawking mode $\hat a_H$.  
This mode, $\phi_D$, is the mode that the detector absorbs when it is excited. 
This mode, by the above argument, has a partner mode, $\phi_P$ which is
orthogonal to $\phi_D$ but is perfectly entangled with $\phi_D$ in the
vacuum state $\ket{0}$. % (defined by $a_k\ket{0}=0$).
If one has a detector to measure various attributes of that partner mode, one
will get vacuum values if one ignores the outcome of the measurements of the
detector mode $\phi_D$. 
Since $\hat a_P$ is a mixture of positive and negative (pseudo) norm vacuum 
modes, one will find a non-zero probability of finding it in the vacuum. 
But that probability would be the same as the probability if one looked into 
the vacuum without measurement of the Hawking/detector mode $\phi_D$. 
There would of course be correlations between the $\hat a_P$ and $\hat a_D$ 
modes. 
I.e., if the Hawking mode were detected (the detector was found in its 
excited state), the detector measuring the partner would also be excited.

Inserting the usual representation of a massless scalar field in 1+1 
dimensional flat space-time 
\bea
\hat \Phi(t,x)= \int {dk\over \sqrt{4\pi|k|}}
\left[\hat a_k e^{-i(|k|t -kx)}+\hat a^\dagger_k e^{i(|k|t -kx)}\right]
,\,
\eea
we have
\bea
\hat a_D 
= 
\frac{\cal N}{i} 
\int d\tau\,\epsilon(\tau)  
\int dk\,\sqrt{|k|\over 4\pi} 
\left(\frac{dt}{d\tau}-\frac{k}{|k|}\,\frac{dx}{d\tau}\right) 
\times
\nn
%&&
\left(
\hat a_k e^{-i(|k|[t(\tau)-x(\tau)]+E\tau)}
%\right.
%\nn
%&&
%\left.
-\hat a_k^\dagger e^{i(|k|[t(\tau)-x(\tau)]-E\tau)}\right)
\,.
\eea
This allows us to read off the Bogoliubov coefficients of 
section~\ref{Definition-Partner} expressing $\hat a_D$, 
the equivalent of the Hawking mode annihilation operator, 
in terms of the operators $\hat a_k$ and $\hat a^\dagger_k$ 
\bea
\alpha_k^*
&=& 
\frac{\cal N}{i} 
\int d\tau\,\epsilon(\tau)\sqrt{|k|\over 4\pi} 
\left(\frac{dt}{d\tau}-\frac{k}{|k|}\,\frac{dx}{d\tau}\right) 
\times
\nn
&&
%{\cal N}\int d\tau\,\epsilon(\tau)\sqrt{|k|\over{2\pi}}\,
e^{-i(|k|[t(\tau)-x(\tau)]+E\tau)}
\,,
\\
\beta_k 
&=& 
i{\cal N}
\int d\tau\,\epsilon(\tau)\sqrt{|k|\over 4\pi} 
\left(\frac{dt}{d\tau}-\frac{k}{|k|}\,\frac{dx}{d\tau}\right) 
\times
\nn
&&
%{\cal N}\int d\tau\,\epsilon(\tau)\sqrt{|k|\over{2\pi}}\,
e^{i(|k|[t(\tau)-x(\tau)]-E\tau)}
\,.
\eea
In general, both the $\alpha$ coefficients and $\beta$ coefficients will be 
non-zero, and $\alpha_k$ will not be proportional to $\beta_k$. 
Thus there is a partner mode. 

Let us now restrict attention to the case where the detector is at rest at 
$x=0$, but $\epsilon(t)$ is non-trivial.
We choose $\epsilon(t)$ such that 
\bea
\scalar{\f{\alpha}}{\f{\beta}}=\int dk\,\alpha^*_{k}\beta_{k}=0
\,,
\ea
so that the detector mode corresponds to pure two-mode squeezing. 
Defining
\bea
\cosh^2r &=& \scalar{\f{\alpha}}{\f{\alpha}} = \int dk\, |\alpha_k|^2 
\,,
\nn
\sinh^2r &=& \scalar{\f{\beta}}{\f{\beta}} = \int dk\, |\beta_k|^2
\,,
\eea
the partner mode is
\bea
\phi_P 
&=& 
\int {dk\over\sqrt{4\pi|k|}}
\left(
\beta_k e^{-i(|k|t-kx)}\coth r 
+
\right.
\nn
&&
\left.
+
\alpha^*_k e^{i(|k|t-kx)}\tanh r
\right) 
\,.
\eea
Now, the term in the first line is just the positive frequency part of 
$-\phi_D^*$, which we can write as 
\bea
\int{dk\over\sqrt{4\pi|k|}}\,\beta_k e^{-i|k|t} 
= 
-\int dt'\,
{\phi_D^*(t',x=0)\over 2\pi i(t-t'-i0^+)}
\,.
\eea
Hence we can also write the partner mode as
\bea
\phi_P(t,x=0)
&=&  
-\coth r
\int dt'\,{\phi_D^*(t',x=0)\over 2\pi i(t-t'-i0^+)}+
\nn
&&
+\tanh r
\int dt'\,{\phi_D^*(t',x=0)\over 2\pi i( t-t'+i0^+)} 
\,. 
\eea
Thus in general $\phi_P(t,0)$ will have a long tail falling off as 
$1/t$ for large $|t|$. 
Only if the moments $\int dt\,t^n \phi_D(t)$ are zero for all $n$ will the
partner fall off faster than any power. 
(If those are zero for all $n<N$ but non-zero thereafter, then $\phi_D$ 
will fall off as $1/|t|^{N+1}$.)

Let us now give an example. Let us assume that 
\bea
\epsilon(t)= \epsilon_0(t) +\epsilon_0(t-T)\lambda\cos[3E(t-T)]
\,,
\eea
where $\lambda$ is very small. 
Again $E$ is the energy difference between the two states of the detector.  
The remaining function $\epsilon_0(t)$ is supposed to be a smooth switching 
function. 
Furthermore let us assume that the Fourier transform of $\epsilon_0(t)$, 
namely $\tilde\epsilon_0(\omega)$ is real and non-zero only in a compact 
region $-E/4<\omega< E/4$. 
The Fourier transform of $\epsilon(t)e^{iEt}$, which occurs in the expression 
for $\hat a_D$, will have three peaks, one small one centered at $-2E$, 
one large one at $E$ and another small one at $4E$. 
Thus $\beta_k$ will have two small peaks with amplitude proportional to 
$\lambda$ at $k\approx\pm2E$, while $\alpha_k$ will have two large peaks 
of amplitude $\ord(1)$ at $k\approx\pm E$ and two of amplitude $\lambda$ at 
$k\approx\pm3E$. 
Because of the limited width of each of these peaks, none overlap, and  
$\f{\beta}$ will be orthogonal to $\f{\alpha}$.
Thus the ``detector mode'' will be a pure two mode squeezed state. 
The $\f{\beta}$-coefficient and thus $\tanh r$ will be of order $\lambda$
while $\coth r$ will be of order $1/\lambda$.
The partner mode will have a temporal Fourier transform with a single peak 
centered at $2E$ of amplitude $\ord(1)$, two smaller peaks of amplitude
$\lambda$ at $-E$, and one of amplitude $\ord(\lambda^2)$ at $-3 E$. 
Thus the partner mode will be approximately given by
\bea
\phi_P(t,x=0)
&=& 
\epsilon_0(t-T)e^{2i E (t-T)}\times\ord(1) 
\nn
&&
+
\epsilon_0(t) e^{-i E t}\times\ord(\lambda)
\nn
&&
+
\epsilon_0(t-T) e^{-3i E (t-T)}\times\ord(\lambda^2)
\,.
\eea
The envelope of the partner mode will thus be dominated by $\epsilon_0(t-T)$, 
i.e., displaced from the detector mode by a time $T$. 
As a result, the partner mode will be centered around a time arbitrarily 
displaced from the maximum of the detector mode. 
Of course, this is somewhat misleading since the part 
$\propto\epsilon_0(t-T)$ of detector mode which leads to detection 
(the $\beta^*$ part of the detector mode) and the $\alpha^*$ 
part of the partner do overlap.  

However, if one chooses some other mode, $\phi_X$ which has an overlap with 
the partner but, let us assume, none with the detector mode, there will be 
correlations between measurements made on this mode and the outcomes of the 
detector measurements.

We note that this example shows that partner modes are not a unique feature of
black holes, or accelerated detectors. 
All detectors, which have a finite probability of detecting something in the 
state of interest, even if due to ``switch on/off'' transients,  will have 
both a ``detector mode'' and a partner mode associated with them. 
If the partner is well separated from the Hawking mode (which it is if we are
interested in the detection of radiation from say a black hole, where the
partner is behind the horizon, and the other is far from the black hole), 
then any measurements made on the partner mode will give results 
indistinguishable from the results in that vacuum state. 
There will however be correlations between the results for measurements on the 
partner and on the ``Hawking mode'' and not with any other modes orthogonal 
to these two. 
If one were able to communicate between the detectors, one could for example
measure the partner whenever the detector detected a particle. 
This would absorb a particle from the vacuum, leaving the vacuum in a lower 
energy state, thus extracting energy from the vacuum -- a form of energy 
teleportation. 
In our case, the ability to communicate the result to somewhere where the 
partner could be detected would be difficult (due to causality) but in some 
cases \cite {hotta} one can actually carry out such a procedure and extract 
energy from the vacuum state leaving the system with locally less energy than 
the vacuum (but leaving the system as a whole of course with higher energy).

\begin{figure}[ht]
\includegraphics[height=8cm]{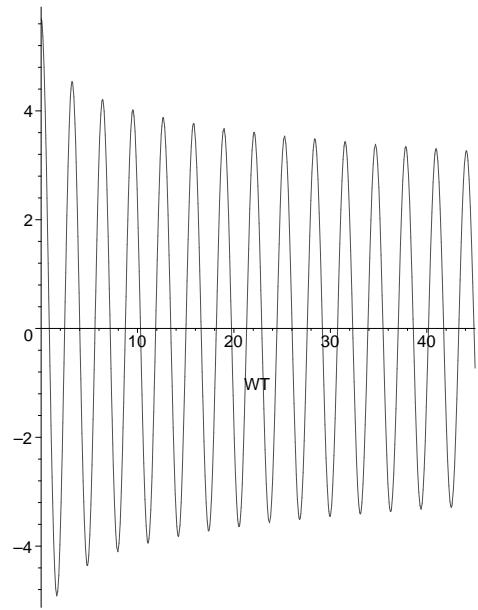}
\caption{\label{figure1} 
The evaluation of $\scalar{\f{\alpha}}{\f{\beta}}(E)$ for the trapezoidal 
coupling as a function of $ET$ with $\tau=1.2~T$. 
The zeros correspond to orthogonality and pure two mode squeezing.}
\end{figure}

Let us also look at a more complex example. 
In this case let us assume that the function $\epsilon(t)$ has the form of a 
trapezoid -- it rises linearly from 0 at time $-\tau$ to $\epsilon_0$ at time  
$-T$, remains constant to time $T$ and then falls linearly to zero at time 
$\tau$.
The Fourier transform of this $\epsilon(t)$ is
\bea
\tilde\epsilon(\omega)
=  
2\epsilon_0
{\cos(\omega T)-\cos(\omega\tau)\over\omega^2(\tau-T)}
\,,
\eea
%
%%% Note sure of the overall factor for this%%%%
%
which gives %{\bf pre-factor...}
\bea
\alpha_k &=& \sqrt{|k|\over4\pi}\,\epsilon_0\, 
{\cos[(|k|- E)T]-\cos[(|k|- E)\tau]\over(|k|- E)^2}
\,,
\nn
\beta_k &=& \sqrt{|k|\over 4\pi}\, \epsilon_0\,
{\cos[(|k|+ E)T]-\cos[(|k|+ E]\tau)\over(|k|+ E)^2}
\,.\,
\eea
The requirement that $\f{\alpha}$ and $\f{\beta}$ be orthogonal can always 
be satisfied for suitable values of $E$. 
Their overlap 
\bea
\scalar{\f{\alpha}}{\f{\beta}}(E)
=
\int dk\,\alpha^*_{k}\beta_{k}
=
2\int\limits_0^\infty dk\,\alpha^*_{k}\beta_{k}
\,,
\eea
is plotted in Figure~\ref{figure1} for $\tau=1.2~T$ as a function of $ET$ 
and we see that there are values of $ET$ which make this overlap zero. 
This is certainly not required, as partners exist even if one does not have 
a pure two mode squeezed state, but it makes, as we saw above, the finding of 
the partner much easier.  
We will use the zero of $\scalar{\f{\alpha}}{\f{\beta}}(E)$
for $E$ nearest 40, namely $E=38.48966$.

\begin{figure}[ht]
\includegraphics[height=8cm]{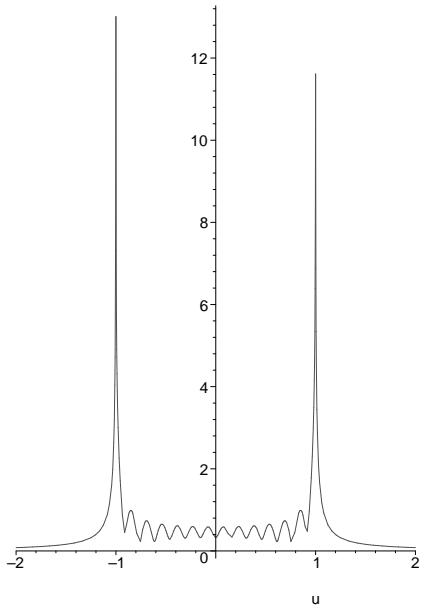}
\caption{\label{figure2} The amplitude $\phi_P$ as a function of $t$.}
\end{figure}

For those  values of $E$ where $\scalar{\f{\alpha}}{\f{\beta}}(E)$ is zero, 
and thus $\f{\alpha}$ is orthogonal to $\f{\beta}$, the partner mode will be 
\bea
\phi_P(t)
= 
\int\limits_0^\infty d\omega
\left( 
C_1\,{\cos[(\omega+ E)\tau]-\cos[(\omega+E)T]\over(\omega+ E)^2}\,e^{i\omega t} 
\right.
\nn
\left.
+C_2\,{\cos[(\omega- E)\tau]-\cos[(\omega- E)T]\over(\omega- E)^2}\,
e^{-i\omega t}
\right]
\,,
\eea
where $C_1$ and $C_2$ are appropriate normalization factors. 
For example, if we take 
\bea
c_a
&=& 
\int d\omega
\left({\cos[(\omega- E)\tau]-\cos[(\omega- E)T]\over(\omega- E)^2}\right)^2
\omega
\nn
&=&
5.15932
\,,
\nn
c_b
&=& 
\int d\omega
\left({\cos[(\omega+ E)\tau]-\cos[(\omega+ E)T]\over(\omega+ E)^2}\right)^2
\omega
\nn
&=&
0.0001195
\,,
\eea
then, defining $\tanh^2r=c_b/c_a$, we get  
\bea
C_1 &=& {\cosh r\over\sqrt{c_b}} = 91.47
\,,
\nn
C_2 &=& {\sinh r\over\sqrt{c_a}} = 1.020\cdot 10^{-5}
\,.
\eea
In this case, the Fourier transform of the partner mode does not vanish at 
$\omega=0$ (because $\tilde\epsilon(E) $ is not zero), but has a step at 
$\omega=0$. 
This implies that the partner mode $\phi_P$ will have a slow falloff of order 
$1/|t|$ for large values of $t$.
In Figure~\ref{figure2} we have a plot of the magnitude of the partner mode 
as a function of $t$ for $T=1,~\tau=1.2$ and $E\approx 40$.
In this case, the partner mode is concentrated in the same area as is the 
original detector mode but with a far longer tail.
However, as we saw above, there is no requirement that the partner mode be
near the peak in the  detector  mode.

%%%%%%%%%%%%%%%%%%%%%%%%%%%%%%%%%%%%%%%%%%%%%%%%%%%%%%%%%%%%%%%%%%%%%%%%%%%%%%
\section{Partners and amplifiers}\label{Partners and amplifiers}
%%%%%%%%%%%%%%%%%%%%%%%%%%%%%%%%%%%%%%%%%%%%%%%%%%%%%%%%%%%%%%%%%%%%%%%%%%%%%%

An example of a system where the detector or Hawking mode is completely 
separate from the partner more is the case of a phase insensitive amplifier. 
Let us take the model of such an amplifier as given in \cite{amplify}, 
in which the amplifier is represented as the coupling, by a free single 
degree of freedom $q$, of two massless one-dimensional fields $\phi$ and 
$\psi$, one ($\psi$) having a negative action
\bea
{\cal L}
&=&
{\cal L}_\phi+{\cal L}_\psi+{\cal L}_q+{\cal L}_{\rm int}
\nn
&=&
\frac12\left[(\partial_t\phi)^2-(\partial_x\phi)^2 \right] 
-\frac12\left[(\partial_t\psi)^2-(\partial_x\psi)^2 \right]+ 
\nn
&&
+\frac12\left[(\partial_t q)^2
+2q(\mu\dot\phi+\nu\dot\psi)
\right]\delta(x)
\,.
\eea
These fields are supposed to live on the positive $x$-axis with Neumann 
boundary conditions at the endpoint.
In order to avoid that the support of the $\delta(x)$-coupling coincides 
with this endpoint, we assume that $x\in[-\varepsilon,\infty)$ and consider 
the limit $\varepsilon\downarrow0$.
The boundary consitions then read 
$\partial_x\phi(t,-\varepsilon)=\partial_x\psi(t,-\varepsilon)=0$. 

This model has solutions %(for $x>0$) 
\bea
\label{eom-phi}
\phi(t,x)&=&\phi_0(t,x) - \mu  q(t-x)
\,, 
\\
\label{eom-psi}
\psi(t,x)&=&\psi_0(t,x) +\nu q(t-x) 
\,,
\\
\partial_t^2 q +(\mu^2-\nu^2)\partial_t q
&=& 
\partial_t[\mu\phi_0(t,0)+\nu\psi_0(t,0)]
\,,
\eea
where $\phi_0$ and $\psi_0$ are solutions to the free (homogeneous)
equation ($\mu=\nu=0$) with the above boundary conditions.
Note that $q$ is damped as long as $\nu^2<\mu^2$.

The quantum operators $\hat\Phi$, $\hat\Psi$, and $\hat Q$ obey the same 
equations.
Taking the Fourier transform and expressing the solutions in terms of
annihilation and creation operators, we have
\bea
\label{free-phi}
\hat\Phi_{0}(t,x)
&=&
\int\limits_0^\infty 
\frac{d\omega}{\sqrt{\pi\omega}}\,
\left(\hat a_\omega e^{-i\omega t}
+\hat a^\dagger_\omega e^{i\omega t}\right)
\cos(\omega x)
\,,
\\
\label{free-psi}
\hat\Psi_{0}(t,x)
&=&
\int\limits_0^\infty 
\frac{d\omega}{\sqrt{\pi\omega}}\,
\left(
\hat b^\dagger_\omega e^{-i\omega t}
+\hat b_\omega e^{i\omega t}\right)
\cos(\omega x)
\,,
\\
\hat Q(t)
&=& 
\hat q_\omega e^{-i\omega t} +\hat q^\dagger_\omega e^{i\omega t}
\,,
\\
\hat q_\omega 
&=& 
\frac{2i}{\sqrt{2\pi}}\,
\frac{\mu\hat a_\omega+\nu\hat b_\omega^\dagger}{\omega+i(\mu^2-\nu^2)}
\,.
\eea
Note that the positions of the creation and annihilation operators of the 
$\hat\Psi$ field is reversed since its conjugate momentum is 
$\pi_\psi=-\partial_t\psi$.
As a result, the inner products of the two fields 
[see also Eq.~(\ref{inner})]
\bea
\inner{\phi}{\tilde\phi}
&=&
i\int\limits_0^\infty dx
\left(\phi^*\tilde\pi_\phi^*-\pi_\phi\tilde\phi^*
\right) 
\,,
\nn
\inner{\psi}{\tilde\psi}
&=&
i\int\limits_0^\infty dx
\left(\psi^*\tilde\pi_\psi^*-\pi_\psi\tilde\psi^*
\right) 
\,,
\ea
are of opposite sign for modes $\phi_\omega$ and $\psi_\omega$ with the 
same $\omega$.
The vacuum state for the $\hat\Psi_0$ field is a maximum of the energy, 
rather than a minimum, and is annihilated by the $\hat b_\omega$ operators.

The initial (input) fields are those that behave as $\phi_{\rm in}(t+x)$ or 
$\psi_{\rm in}(t+x)$
while the final (output) fields go as $\phi_{\rm out}(t-x)$ or 
$\psi_{\rm out}(t-x)$, respectively (remember that $x>0$). 
Assuming that the input fields are the free fields in Eqs.~(\ref{free-phi})
and (\ref{free-psi}), 
\bea
\hat\Phi_{\rm in}
=
\hat\Phi^{\rm in}_0 
= 
\int\limits_0^\infty 
\frac{d\omega}{\sqrt{4\pi\omega}}\,
\left(\hat a_\omega e^{-i\omega(t+x)}+\hat a^\dagger e^{i\omega (t+x)}\right) 
\,,
\eea
and similarly for $\hat\Psi_{\rm in}=\hat\Psi^{\rm in}_0$, the output part is,
according to Eqs.~(\ref{eom-phi}) and (\ref{eom-psi}), given by
\bea
\hat\Phi_{\rm out}(t-x)  
&=& 
\int\limits_0^\infty
\frac{d\omega}{\sqrt{4\pi\omega}}\,
e^{-i\omega (t-x)}
\left(\hat a_\omega\,\frac{\omega-i(\mu^2+\nu^2)}{\omega+i(\mu^2-\nu^2)}
\right.
\nn
&&
\left.
- 
\hat b^\dagger_\omega 
\frac{2i\mu\nu}{\omega+i(\mu^2-\nu^2)}
\right) 
+{\rm h.c.,}
\\
\hat\Psi_{\rm out}(t-x)  
&=& 
\int\limits_0^\infty
\frac{d\omega}{\sqrt{4\pi\omega}}\,
e^{-i\omega (t-x)}
\left(\hat b^\dagger_\omega \,
\frac{\omega-i(\mu^2+\nu^2)}{\omega-i(\mu^2-\nu^2)}
\right.
\nn
&&
\left.
+\hat a_\omega
\frac{2i\mu\nu}{\omega-i(\mu^2-\nu^2)}
\right) 
+{\rm h.c.,}
\eea
where $\rm h.c.$ is the Hermitian conjugate. 
Thus, one can write the output annihilation and creation operators in terms 
of the input by
\bea
\hat A_\omega
&=& 
\hat a_\omega\,\frac{\omega-i(\mu^2+\nu^2)}{\omega+i(\mu^2-\nu^2)}
- 
\hat b^\dagger_\omega 
\frac{2i\mu\nu}{\omega+i(\mu^2-\nu^2)}
\,,
%
%{-i\omega -(\mu^2+\nu^2)\over -i\omega +(\mu^2-\nu^2)} a_\omega -
%{2\mu\nu\over -i\omega +(\mu^2-\nu^2)} b^\dagger_\omega
\\
\hat B^\dagger_\omega 
&=& 
\hat b^\dagger_\omega \,
\frac{\omega-i(\mu^2+\nu^2)}{\omega-i(\mu^2-\nu^2)}
+\hat a_\omega
\frac{2i\mu\nu}{\omega-i(\mu^2-\nu^2)}
\,.
%{-i\omega +(\mu^2+\nu^2)\over -i\omega
%+(\mu^2-\nu^2)}b^\dagger_\omega + {2\mu\nu\over  -i\omega 
%+(\mu^2-\nu^2)} a_\omega
\eea
Writing the first equation as 
$\hat A_\omega=\alpha_\omega\hat a_\omega+\beta_\omega\hat b^\dagger_\omega$ 
we see that the factor $|\alpha_\omega|$ is larger than unity 
(unless $\nu=0$) which means that signals in the $\hat a_\omega$ channel 
are amplified. 
However, as required by unitarity, this goes along with additional noise 
stemming from the $\hat b^\dagger_\omega$ term. 

Now let us consider a mode in the $\phi$ output channel, say, defined by at
late times 
\bea
f_H(t-x)
= 
\int\frac{d\omega}{\sqrt{4\pi\omega}}\,
\tilde f_H(\omega)\,e^{-i\omega (t-x)}
\,,
\eea
where we will assume that $\tilde f_H^\omega$ is non-zero only for 
$\omega>0$  and is normalized so that 
$\int d\omega|\tilde f_H(\omega)|^2=1$. 
The operator associated with this mode at late times is 
%($t\rightarrow \infty$) is 
%
\bea
\hat a_H
&=&
\inner{f_H}{\hat\Phi}
=
\int d\omega\,\tilde f_H^*(\omega)\hat A_\omega
=
\int\limits_0^\infty d\omega\,\tilde f_H^*(\omega) 
\times
\nn
&&
\left(
\hat a_\omega\,\frac{\omega-i(\mu^2+\nu^2)}{\omega+i(\mu^2-\nu^2)}
- 
\hat b^\dagger_\omega 
\frac{2i\mu\nu}{\omega+i(\mu^2-\nu^2)}
\right).  
\eea
(We use the $\hat{}$ to remind ourselves that this operator may be in either
of the two channels, before or after the the interaction with $q$)
As a result, the partner mode will be
\bea
\hat a_P^\dagger 
&=&
\int\limits_0^\infty d\omega\,\tilde f_H^*(\omega) 
\left( 
\hat a_\omega\,
\frac{\omega-i(\mu^2+\nu^2)}{\omega+i(\mu^2-\nu^2)}\,
\tanh\vartheta
\right. 
\nn
&&
\left.
-
\hat b^\dagger_\omega
\frac{2i\mu\nu}{\omega+i(\mu^2-\nu^2)}\,\coth\vartheta
\right) 
\,,
\eea
where the mixing angle is given by 
\bea
\sinh^2\vartheta= 
\int\limits_0^\infty d\omega\,\left|\tilde f_H^*(\omega)\right|^2
{4\mu^2\nu^2\over \omega^2+(\mu^2-\nu^2)^2} 
\,.
\eea

If we define the following frequency dependent mixing angles and phases 
\bea
\cosh\theta_\omega&=& \left\vert {-i\omega-(\mu^2+\nu^2)
\over -i\omega+(\mu^2-\nu^2)}\right\vert
\,,
\\
\sinh\theta_\omega&=& \left\vert {2\mu\nu\over
i\omega+(\mu^2-\nu^2)}\right\vert
\,,
\\
e^{i\sigma_\omega}&=& {-i\omega+(\mu^2-\nu^2)\over |-i\omega+(\mu^2-\nu^2)|}
\,,
\\
e^{i\lambda}&=& {(\mu^2+\nu^2)+i\omega 
\over|i\omega+(\mu^2+\nu^2)|}
\,,
\eea
then 
\bea
\sinh^2\vartheta = 
\int\limits_0^\infty d\omega\,\left|\tilde f_H^*(\omega)\right|^2
%\int |\kappa_\omega|^2 
\sinh^2\theta_\omega
% d\omega
\,.
\eea

Then we find that we can express the annihilation operator of the partner mode
in terms of the outgoing creation and annihilation operators in terms of
either the input annihilation operators, or of the output
\bea
a_P^\dagger&=& 
\int\limits_0^\infty d\omega\,\tilde f_H^*(\omega) 
%\int d\omega \kappa^*_\omega 
e^{-i\sigma_\omega} 
\left (- \tanh(\vartheta)
\cosh(\theta_\omega) e^{i\lambda_\omega} \hat a_\omega 
\right.
\nn
&&
\left.
- \coth(\vartheta)
\sinh(\theta_\omega) \hat b_\omega^\dagger\right)\\
&=& 
\int\limits_0^\infty d\omega\,\tilde f_H^*(\omega) 
%\int d\omega \kappa_\omega^* 
e^{-i\sigma_\omega} \left(2{ \sinh^2\vartheta-\sinh^2\theta_\omega\over
\sinh(2\vartheta) } \hat A_\omega 
\right.
\nn
&&
\left.
- {\sinh(2\theta_\omega)\over
\sinh(2\vartheta)} \hat B_\omega^\dagger\right)
\eea

If $|\tilde f_H^*(\omega)|$ is a highly peaked function about the frequency 
$\omega$ such that $\theta_\omega=\vartheta$, then the first
term will be zero and the partner mode, made up entirely of $B_\omega$, will be
confined completely to the second output channel-- the $\psi$ channel.
However, if $\kappa$ is a broad function 
(non zero over a range of order or  larger than
$\mu^2-\nu^2$, then the partner mode will have support in both the $\psi$ and
the $\phi$ output channels -- mostly in the former, but partially in the latter
as well. 

This will also be true in the Black hole case as well, which behaves exactly
like this amplifier, with the output $\phi$ and $\psi$ channels being the
modes travelling to infinity and those falling into the singularity
respectively. For highly peaked functions of frequency, the partner is behind
the horizon, while for broadly peaked functions of frequency, the partner has
components both inside and outside the horizon. This is another indication of
the non-linear nature of the partner mode. Since any mode is the sum of highly
peaked functions, one might expect that the a broadly peaked Hawking mode
might still have a partner entirely behind the horizon, but it does not.

%%%%%%%%%%%%%%%%%%%%%%%%%%%%%%%%%%%%%%%%%%%%%%%%%%%%%%%%%%%%%%%%%%%%%%%%%%%%%%
\section{Moving Mirror Radiation}\label{Moving Mirror Radiation}
%%%%%%%%%%%%%%%%%%%%%%%%%%%%%%%%%%%%%%%%%%%%%%%%%%%%%%%%%%%%%%%%%%%%%%%%%%%%%%

Before applying the concept of partner particles to what has been taken to be 
a simple toy model for black hole evaporation -- the radiation given off by 
an exponentially accelerated mirror 
\cite{Davies+Fulling,Walker,Carlitz+Willey,wilczek} -- 
let us briefly review the basic concepts. 
We consider a massless scalar field in 1+1 dimensional flat space-time 
%($\hbar=c=1$)
%
\bea
\Box\phi=0
\,.
\ea
At a point-like mirror with the trajectory $x_{\rm m}(t)$, we impose 
Dirichlet boundary condition 
\bea
\label{Dirichlet}
\phi(t,x_{\rm m}[t])=0
\,.
\ea
In terms of the light-cone coordinates 
\bea
u=t-x
\,,\quad
v=t+x
\,,
\ea
the general solution of $\Box\phi=0$ without the boundary 
condition~(\ref{Dirichlet}) can be written as a sum of independent 
left-moving $\phi_{\rm left}(v)$ and right-moving $\phi_{\rm right}(u)$
contributions 
$\phi(u,v)=\phi_{\rm left}(v)+\phi_{\rm right}(u)$.
The boundary condition~(\ref{Dirichlet}) imposes constraints 
on these two parts and thus the quantum field can be decomposed as 
\bea
\label{quantum-field}
\hat\phi(u,v)=
\int\limits_0^\infty
d\omega\,
\frac{e^{-i\omega v}-e^{-i\omega(2\tau[u]-u)}}{\sqrt{4\pi\omega}}\,
\hat a_\omega^{\rm in}
+{\rm h.c.}
\ea
Here  $(\hat a_\omega^{\rm in})^\dagger$ and $\hat a_\omega^{\rm in}$ 
denote the initial creation and annihilation operators and the function 
$\tau[u]$ is implicitly determined by the mirror trajectory 
\bea
\tau[u]=u+x_{\rm m}(\tau[u]) 
\,.
\ea
Hence the mode functions in Eq.~(\ref{quantum-field}) automatically satisfy 
the boundary condition~(\ref{Dirichlet}). 
For a mirror at rest $x_{\rm m}=\rm const$, we find $\tau[u]=u+x_{\rm m}$ 
and thus these mode functions simplify to 
$e^{-i\omega v}-e^{-i\omega(u+2x_{\rm m})}$ which just gives  
$2ie^{-i\omega t}\sin(\omega[x_{\rm m}-x])e^{-i\omega x_{\rm m}}$ 
as one would expect. 
Thus, assuming that the mirror is a rest initially, the initial vacuum state 
is determined by 
\bea
\label{vacuum}
\forall_{\omega>0}\;
\hat a_\omega^{\rm in}\ket{0}=0
\,.
\ea
Similarly, for a mirror $x_{\rm m}=Vt$ moving with a constant velocity $V$,
we get $\tau[u]=u/(1-V)$.
In these cases, no particles are created -- but with an accelerated motion 
of the mirror (resulting in a non-trivial form of $\tau[u]$) 
one can create particles out of the initial vacuum. 

In terms of the light-cone coordinates, the proper acceleration of the mirror 
$\ddot x_{\rm m}/(1-\dot x_{\rm m}^2)^{3/2}$ can be written as 
$\ddot v_{\rm m}/(\dot u_{\rm m}\dot v_{\rm m})^{3/2}$.
Similarly, the red-shift factor $\sqrt{(1+\dot x_{\rm m})/(1-\dot x_{\rm m})}$ 
simply reads $\sqrt{\dot v_{\rm m}/\dot u_{\rm m}}$.
Now, if we choose the mirror trajectory in such a way that the proper 
acceleration 
of the mirror is proportional to the red-shift factor, an observer at rest sees 
a stationary thermal spectrum given off by the moving mirror.
This situation corresponds to the mirror trajectory 
\bea
\label{trajectory}
t+x_{\rm m}
=
v_{\rm m}
=
-\frac{e^{-\kappa u_{\rm m}}}{\kappa}
=
-\frac{e^{-\kappa (t-x_{\rm m})}}{\kappa}
\,,
\ea
where $\kappa$ is a proportionality constant which sets the temperature.
As a result, the mode functions satisfying the boundary 
condition~(\ref{Dirichlet})
are given by 
\bea
\phi_\omega(u,v)
=
e^{-i\omega v}-\exp\left\{i\,\frac{\omega}{\kappa}\,e^{-\kappa u}\right\}
\,.
\ea
In principle, since the proper acceleration of the 
trajectory~(\ref{trajectory}) 
vanishes for very early times $t\downarrow-\infty$, we could consider a mirror 
moving along the world-line~(\ref{trajectory}) for all times. 
However, to make the initial behavior as simple as possible, we assume that 
the mirror is initially at rest and starts accelerating along the 
trajectory~(\ref{trajectory}) 
at $u^0=0$ which means $v_{\rm m}^0=-1/\kappa$, i.e., 
\bea
v_{\rm m}(u)
= 
-\frac{1}{\kappa}\,
\left\{
\begin{array}{lll}
1-\kappa u & {\rm for} & u<0 \\
e^{-\kappa u}
& {\rm for} & u>0
\end{array}
\right.
\,.
\ea
Consequently, incoming light rays with $v<-1/\kappa$ are reflected by the 
mirror at rest, i.e., initial waves of the form $e^{-i\omega v}$ are simply 
transformed to final waves of the form $e^{-i\omega u}$. 
Incoming light rays in the window $-1/\kappa<v<0$ are reflected by the 
accelerating mirror.
In this region, initial waves of the form $e^{-i\omega v}$ are stretched by the 
increasing red-shift factor and finally behave as 
$\exp\{i\omega e^{-\kappa u}/\kappa\}$.
More generally, an initial wave-packet of the form $\phi_{\rm left}(v)$ in the 
region $-1/\kappa<v<0$ is transformed to $\phi_{\rm right}(-e^{-\kappa u}/\kappa)$,
i.e., a final right-moving wave-packet in the region $u>0$. 
The remaining light rays with $v>0$ do not see the mirror at all and thus their 
functional form is unchanged. 
Hence the null line $v=0$ is analogous to the black hole horizon.

Starting in the initial vacuum state~(\ref{vacuum}), we can derive the 
two-point functions in the final state. 
In order to avoid artifacts %get rid of the undetermined constant 
stemming from the infra-red divergence of the massless scalar field in two 
dimensions, we consider the first derivatives of the fields.
(This is somewhat similar to considering the electric and magnetic fields 
instead of the scalar and vector potentials.) 
As mentioned above, the field can be split up into a left-moving 
$\hat\phi_{\rm left}(v)$ and a right-moving part $\hat\phi_{\rm right}(u)$.
The correlation between two final the left-moving contributions 
(with $v_{1,2}>0$) gives 
\bea
\bra{0}
\partial_v\hat\phi_{\rm left}(v_1)\partial_v\hat\phi_{\rm left}(v_2)
\ket{0}
=
-\frac{1}{4\pi}\,\frac{1}{(v_1-v_2)^2}
\,,
\ea
which just reflects the fact the associated quantum state is locally 
indistinguishable from vacuum (since it has not ``seen'' the mirror at all).  

Considering the correlation between two right-moving contributions which have 
been  reflected by the accelerated mirror (with $u_{1,2}>0$), however, gives  
\bea
\bra{0}
\partial_u\hat\phi_{\rm right}(u_1)\partial_u\hat\phi_{\rm right}(u_2)
\ket{0}
=
%\times
\nn
%\times
-\frac{\kappa^2}{16\pi}
\frac{1}{\sinh^2(\kappa[u_1-u_2]/2)}
\,.
\ea
As already suggested by the periodicity in imaginary time (KMS condition), 
this is locally indistinguishable from a thermal state with the temperature 
\bea
T_H=\frac{\kappa}{2\pi}
\,.
\ea
The fact that this thermal contribution and the above vacuum part are actually 
just two regions of the same pure state results in non-trivial 
cross-correlations between these two modes 
\bea
\bra{0}
\partial_v\hat\phi_{\rm left}(v_1)\partial_u\hat\phi_{\rm right}(u_2)
\ket{0}
=
-\frac{\kappa^2}{4\pi}
%\times
%\nn
%\times
\frac{e^{-\kappa u_2}}{(e^{-\kappa u_2}+\kappa v_1)^2}
\,.\;
\ea
These results can be generalized to different mirror trajectories $v_{\rm m}(u)$ 
in a straightforward manner.
In this case, the mode functions read $e^{-i\omega v}-e^{-i\omega v_{\rm m}(u)}$ 
and thus the two-point function is given by 
\bea
\bra{0}\hat\phi(u_1,v_1)\hat\phi(u_2,v_2)\ket{0}
=
%\times
\nn
%\times
-\frac{1}{4\pi}
\,\ln\left(
\frac{[v_1-v_2][v_{\rm m}(u_1)-v_{\rm m}(u_2)]}
{[v_1-v_{\rm m}(u_2)][v_{\rm m}(u_1)-v_2]}
\right)
%
%\frac{(\partial_u v_{\rm m}[u_1])(\partial_u v_{\rm m}[u_2])}
%{(v_{\rm m}[u_1]-v_{\rm m}[u_2])^2}
\,.
\ea
%
%up to a possible additive constant reflecting the aforementioned 
%infra-red divergence.  

%%%%%%%%%%%%%%%%%%%%%%%%%%%%%%%%%%%%%%%%%%%%%%%%%%%%%%%%%%%%%%%%%%%%%%%%%%%%%%
\section{Partner Particles for Mirror thermal radiation}
%%%%%%%%%%%%%%%%%%%%%%%%%%%%%%%%%%%%%%%%%%%%%%%%%%%%%%%%%%%%%%%%%%%%%%%%%%%%%%

Now let us try to determine the partner particles for the thermal 
radiation created by the mirror as an analogue for Hawking radiation. 
Thus, we define the outgoing Hawking wave function $f_H(u)$ as a linear 
combination of final positive-frequency right-moving plane waves 
\bea
\label{Hawking-mode}
f_H(u)=\int\limits_0^\infty d\Omega\,\tilde f_H(\Omega)\, e^{-i\Omega u}
\,,
\ea
where $\tilde f_H(\Omega)$ is then the Fourier transform of $f_H(u)$.
Due to the restriction to positive final frequencies, the support of $f_H(u)$ 
is unbounded, i.e., extends to negative $u$ as well. 
However, for simplicity, we assume that $f_H(u)$ lies mostly in the thermal 
region $u>0$, i.e., that $f_H(u)$ is exponentially suppressed for $u<0$. 
Alternatively, we could consider the case of eternal acceleration of the 
mirror, where the thermal region extends to negative $u$. 

The Bogoliubov coefficients can then defined by the overlap between these 
modes~(\ref{Hawking-mode}) and the initial positive/negative frequency modes  
$e^{\pm i\omega v}$. 
For the trajectory~(\ref{trajectory}),  these overlap integral can be 
calculated analytically in terms of $\Gamma$-functions etc. 
However, instead of using these $\Gamma$-functions, we do the following trick:
Initially, the mode~(\ref{Hawking-mode}) behaved as 
\bea
\label{Hawking-initial}
f_H^{\rm in}(v<0)=\int\limits_0^\infty d\Omega\,\tilde f_H(\Omega)\, 
(-\kappa v)^{i\Omega/\kappa} 
%e^{-i\Omega u}
\,,
\ea
and $f_H^{\rm in}(v>0)=0$.
Now, let us consider the following linear combinations 
\bea
\label{linear-combination}
%\frac{1}{\sqrt{4\sinh(\pi\omega/\kappa)}}
%\times
f^\pm_\Omega(v)
=
\left\{
\begin{array}{lll}
e^{\pm\pi\Omega/(2\kappa)}\,\left|\kappa\,v\right|^{i\Omega/\kappa} 
& {\rm for} & v<0 \\
e^{\mp\pi\Omega/(2\kappa)}\,\left|\kappa\,v\right|^{i\Omega/\kappa} 
& {\rm for} & v>0
\end{array}
\right.
\,.
\ea
These linear combinations are chosen such that the function $f^+_\Omega(v)$ 
is holomorphic in the entire lower half of the complex $v$ plane, i.e., 
for $\Im(v)<0$, and has a singularity at $v=0$ as well as a branch cut 
from $v=0$ to $v=\infty$ in the upper half.
On the other hand, recalling the structure of the initial mode functions 
$e^{-i\omega v}$, we find that any solution is exactly
decomposed of positive (initial) frequency modes if and only if it is 
holomorphic in entire lower half of complex $v$ plane.
Thus the linear combination $f^+_\Omega(v)$ in Eq.~(\ref{linear-combination}) 
contains only positive (initial) frequency modes for all $\Omega$, 
i.e., it corresponds to an initial annihilation operator 
$\hat a_\Omega^{\rm in}$ with $\hat a_\Omega^{\rm in}\ket{0}=0$. 
Conversely, the other combination  $f^-_\Omega(v)$ is holomorphic 
in the upper half of the complex $v$ plane and thus contains negative (initial)
frequencies only, i.e., it corresponds to an initial creation operator.
Using the symmetry $[f^\pm_\Omega(v)]^*=f^\mp_{-\Omega}(v)$,
we find that $f^-_\Omega(v)$ corresponds to 
$(\hat a^{\rm in}_{-\Omega})^\dagger$. 

Now, we can decompose the function $|\kappa v|^{i\Omega/\kappa}$ 
as a linear combination of $f^\pm_\Omega(v)$ which gives 
\bea
\frac{e^{+\pi\Omega/(2\kappa)} f^+_\Omega(v) -  
e^{-\pi\Omega/(2\kappa)}  f^-_\Omega(v)}
{\sqrt{2\sinh(\Omega/\kappa)}}
=
\left|\kappa\,v\right|^{i\Omega/\kappa} 
\,,
\ea
for $v<0$ and vanishes for $v>0$.
This enables us to directly read off the decomposition of the final Hawking 
mode~(\ref{Hawking-mode}) into initial creation and annihilation operators 
\bea
\hat a_H=\int\limits_0^\infty d\Omega\,\tilde f_H(\Omega)\,
\left[
\alpha_\Omega\,\hat a_\Omega^{\rm in}
+
\beta_\Omega\left(\hat a_{-\Omega}^{\rm in}\right)^\dagger
\right]
\,,
\ea
with the Bogoliubov coefficients 
\bea
\alpha_\Omega=\frac{e^{+\pi\Omega/(2\kappa)}}{\sqrt{2\sinh(\Omega/\kappa)}}
\,,\;
\beta_\Omega=\frac{e^{-\pi\Omega/(2\kappa)}}{\sqrt{2\sinh(\Omega/\kappa)}}
\,,
\ea
where the denominator $\sqrt{2\sinh(\Omega/\kappa)}$ ensures the correct 
normalization $|\alpha_\Omega|^2-|\beta_\Omega|^2=1$.

This is the starting point for the derivation of the partner mode $\hat a_P$.
Note that the modes $f^\pm_\Omega(v)$ are orthogonal with respect to 
the usual inner product for the scalar field 
\bea
\label{inner}
\inner{\phi_1}{\phi_2}
=
i\int d\Sigma^\mu\,
\phi_1^*\,\stackrel{\leftrightarrow}{\partial}_\mu\,\phi_2
\,,
\ea
where $\phi_1^*\,\stackrel{\leftrightarrow}{\partial}_\mu\,\phi_2=
\phi_1^*\,\partial_\mu\,\phi_2-\phi_2\,\partial_\mu\,\phi_1^*$. 
For purely right-moving modes, we may align the hyper-surface $\Sigma$ 
with a null-line of constant $u$ (or even ${\cal J}^-$) such that 
the $d\Sigma^\mu$-integral becomes an integration over $v$ and the 
derivative $\stackrel{\leftrightarrow}{\partial}_\mu$ simplifies to 
$\stackrel{\leftrightarrow}{\partial}_v$.
As explained above, the mode functions $f^+_\Omega(v)$ are decomposed of 
purely positive frequency initial waves $e^{-i\omega v}$ with $\omega>0$.
Conversely, the mode functions $f^-_\Omega(v)$ are decomposed of 
purely negative frequency initial waves $e^{+i\omega v}$ with $\omega>0$.
As a result,  the contributions $f^+_\Omega(v)$ and $f^-_\Omega(v)$ 
are orthogonal and thus the vectors $\f{\alpha}$ and $\f{\beta}$ are orthogonal,
i.e., we have pure two-mode squeezing for all $\tilde f_H(\Omega)$.

\begin{figure}[ht]
\includegraphics[height=8cm]{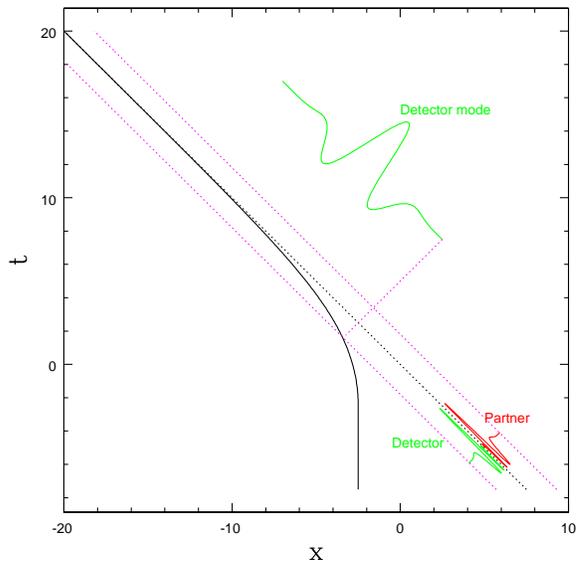}
\caption{\label{figure3} 
The outgoing detected Hawking mode, its shape when traced back to the 
initial state, and its partner mode.}
\end{figure}

Using the arguments presented in Sec.~\ref{Definition-Partner}, 
we find that the partner mode reads 
\bea
\hat a_P
=
\int\limits_0^\infty d\Omega\,\tilde f_H^*(\Omega)\,
\left[
\chi\beta_\Omega\,\hat a_{-\Omega}^{\rm in}
+
\chi^{-1}\alpha_\Omega\left(\hat a_{\Omega}^{\rm in}\right)^\dagger
\right]
\,,
\ea
with the factor $\chi=\alpha/\beta$
\bea
\chi^2
=
\frac{\int\limits_0^\infty d\Omega\,|\tilde f_H(\Omega)|^2 \alpha_\Omega^2}
{\int\limits_0^\infty d\Omega\,|\tilde f_H(\Omega)|^2 \beta_\Omega^2}
\,.
\ea
If the Hawking mode $\tilde f_H(\Omega)$ is well localized and peaked at a 
given frequency $\Omega=\Omega_0$, then we may approximate this factor by 
$\chi\approx e^{ \pi\Omega_0/\kappa}$.
As a result, the wave function of the partner particle $f_P(v)$ contains the 
following linear combination of the modes $f^\pm_\Omega(v)$ which yields  
\bea
\label{cancel}
\frac{e^{+\pi\Omega/(2\kappa)} f^+_{-\Omega}(v) -  
e^{-\pi\Omega/(2\kappa)}  f^-_{-\Omega}(v)}
{\sqrt{2\sinh(\Omega/\kappa)}}
=
\left|\kappa\,v\right|^{-i\Omega/\kappa} 
\,,
\ea
for $v>0$ and vanishes for $v<0$.
Thus the wave-function of the partner particle is approximately the mirror 
image of the initial form of the Hawking mode~(\ref{Hawking-initial}) 
on the other side of the horizon at $v=0$, i.e., 
\bea
\label{partner-mirror}
f_P(v)
\approx 
\int\limits_0^\infty d\Omega\,\tilde f_H^*(\Omega)\, 
(\kappa v)^{-i\Omega/\kappa} 
=
f^*_H\left[-\frac{\ln(\kappa v)}{\kappa}\right] 
\,,
\ea
for $v>0$ and zero for $v<0$. 
As a result, detecting a Hawking particle with, say, $\Omega=\ord(\kappa)$ 
in the thermal region at late times $\kappa u\gg1$ yields (up to normalization)
approximately the same state as creating a partner particle with exponentially 
short wavelengths in the left-moving vacuum region at very small but positive 
values of $v$.
Figure~\ref{figure3} is a plot of of a specific outgoing Hawking mode 
(say the mode detected by some detector) and that mode in the input state, 
and its partner mode.

If we could signal the measurement result of the Hawking detector at large 
$u>0$ on the right-hand side to this vacuum region on the left-hand side, 
we would (at least in principle) be able to extract energy out of this quantum 
state, which is locally indistinguishable from vacuum -- this is directly 
related to the concept of ``energy teleportation''. 
However, causality prevents us from signaling since these two events are 
space-like separated. 

Another point is that the cancellation of the contributions in 
Eq.~(\ref{cancel}) occurs at one frequency $\Omega=\Omega_0$ only.
If we consider a small but finite width $\Delta\Omega\ll\Omega_0$, 
the partner mode will also have support in the thermal region $u>0$, 
i.e., $v<0$. 
For simplicity, let us consider a Gaussian wave-packet of the form 
\bea
\label{Gaussian-wave-packet}
\tilde f_H(\Omega)={\cal N}\exp\left\{
-\frac{(\Omega-\Omega_0)^2}{2(\Delta\Omega)^2}+i\Omega u_0
\right\}
\,,
\ea
which is centered around $\Omega_0$ in frequency space and 
around $u_0$ in position space.
Even though this $\tilde f_H(\Omega)$ is not exactly zero for $\Omega<0$,
this contribution is exponentially small  for $\Delta\Omega\ll\Omega_0$
and thus negligible. 
Similarly, the tail of this wave-packet for $u<0$ can be made very small
by assuming $u_0\Delta\Omega\gg1$.

After inserting this form~(\ref{Gaussian-wave-packet}), 
the formula for the partner mode $f_P(v<0)$ contains an 
$\Omega$-integral whose integrand vanishes (to lowest order) 
at $\Omega=\Omega_0$.
Taylor expanding this integrand around this zero then yields 
a first-order contribution of the form $(\Omega-\Omega_0)\tilde f_H(\Omega)$
which can also be represented by 
$(\Delta\Omega)^2[iu_0-\partial_\Omega]\tilde f_H(\Omega)$.
After the Fourier transformation~(\ref{Hawking-mode}), the $\partial_\Omega$
translates into $iu$ and thus the partner wave function acquires a 
small contribution with the same support as the Hawking mode 
\bea
f_P(u)\sim\frac{(\Delta\Omega)^2}{\kappa}\,[u-u_0]f_H(u)
\,,
\ea
in addition to the dominant contribution~(\ref{partner-mirror}).
Due to the term $[u-u_0]$, the two modes are orthogonal as they should be. 

%%%%%%%%%%%%%%%%%%%%%%%%%%%%%%%%%%%%%%%%%%%%%%%%%%%%%%%%%%%%%%%%%%%%%%%%%%%%%%
%\section{Black Hole Evaporation}
%%%%%%%%%%%%%%%%%%%%%%%%%%%%%%%%%%%%%%%%%%%%%%%%%%%%%%%%%%%%%%%%%%%%%%%%%%%%%%

%mirror falling into a black hole 

%partner particles mostly in locally approximate 
%vacuum state falling towards singularity 

%%%%%%%%%%%%%%%%%%%%%%%%%%%%%%%%%%%%%%%%%%%%%%%%%%%%%%%%%%%%%%%%%%%%%%%%%%%%%%
\section{Conclusions}
%%%%%%%%%%%%%%%%%%%%%%%%%%%%%%%%%%%%%%%%%%%%%%%%%%%%%%%%%%%%%%%%%%%%%%%%%%%%%%

Perhaps the most surprising conclusion of this paper is that the partner 
particles of the thermal radiation (emitted by a mirror or a black hole) 
are concentrated in a region which is locally indistinguishable from vacuum.
In the black hole evaporation process, the Hawking particles emitted at 
early or intermediate times can be entangled not with some other energetic 
emission at late times, but with final vacuum fluctuations. 
This weakens the usual argument in black hole evaporation studies which assume 
unitarity in the above sense, which states that the large amount of 
information left in the black hole 
(entanglement with the emitted thermal emission) must be accompanied 
by the eventual emission of large amounts of energy.

Thus, if one were to imagine a black hole constantly fed by a pure state 
designed to just compensate for the energy emitted by the black hole in 
Hawking thermal emission, for $10^{99}$ times the natural decay lifetime 
of the black hole, there must be a huge amount of information inside the 
black hole, encoded in the entanglement with the outgoing Hawking radiation. 
When the black hole eventually evaporates that information, which must be 
emitted at late times, one could expect that it must be accompanied by a 
large amount of energy as well. 
However this paper offers the possibility that that eventual emission of 
information could be in the form of the vacuum, and carrying no energy. 

For example, the Bardeen model \cite{bardeen}, a response to the 
AMPS \cite{amps} argument that either unitarity (as defined above) 
or the regularity of the space-time at horizon must be wrong, has the partner 
radiation to the Hawking emission trapped within the apparent horizon of the 
black hole, until eventually that horizon disappears. 
This would seem to require a massive emission of energy just at the time when 
that apparent horizon disappears to accompany that massive emission of 
information (i.e., the entangled partner radiation to the earlier Hawking 
emission). 
Our results offer the possibility that those partner modes are, in that final 
stage, simply a part of the vacuum state with no energy accompanying them.

%%%%%%%%%%%%%%%%%%%%%%%%%%%%%%%%%%%%%%%%%%%%%%%%%%%%%%%%%%%%%%%%%%%%%%%%%%%%%%
\acknowledgments
%%%%%%%%%%%%%%%%%%%%%%%%%%%%%%%%%%%%%%%%%%%%%%%%%%%%%%%%%%%%%%%%%%%%%%%%%%%%%%

The authors acknowledge support from the 
Yukawa Institute for Theoretical Physics (YITP) 
and the YITP Workshop on Quantum Information Physics (YQIP2014).  
W.G.U.\ obtained support from NSERC of Canada, the Templeton foundation, 
and the Canadian Institute for Advanced Research.
R.S.\ acknowledges support from DFG (SFB-TR12). 
We also thank the Perimeter Institute for Theoretical Physics (PI)
for hospitality and support where part of this work was done
and the Banff International Research Station for inviting us all to the 
workshop ``Entanglement in Curved Spacetime'' where this work was begun.

\appendix 

%%%%%%%%%%%%%%%%%%%%%%%%%%%%%%%%%%%%%%%%%%%%%%%%%%%%%%%%%%%%%%%%%%%%%%%%%%%%%%
\section{Uniqueness Proof}
%%%%%%%%%%%%%%%%%%%%%%%%%%%%%%%%%%%%%%%%%%%%%%%%%%%%%%%%%%%%%%%%%%%%%%%%%%%%%%

In the following, we show that the ansatz~(\ref{partner-ansatz}) for the 
partner particle is the most general ansatz one can make, i.e., that the 
partner particle is uniquely determined by our two conditions 
(unless we have pure single-mode squeezing $\f{\alpha}\|\f{\beta}$).
For free fields, the initial vacuum state is a Gaussian state and thus 
the state restricted to the two modes $\hat a_H$ and $\hat a_P$ 
must also be a Gaussian state. 
In the position representation, i.e., as a function of the position vector
$\f{x}=(x_H,x_P)$ where 
$\hat x_H=(\hat a_H+\hat a_H^\dagger)/\sqrt{2}$ and 
$\hat x_P=(\hat a_P+\hat a_P^\dagger)/\sqrt{2}$, 
the most general wave function of a pure Gaussian state reads 
\bea
\psi(\f{x})={\cal N}\exp\left\{-\frac12\,\f{x}\cdot\f{M}\cdot\f{x}\right\}
\,,
\ea
where $\f{M}$ is a symmetric but possibly complex matrix 
and $\cal N$ the corresponding normalization factor.
In order to have a normalizable state, the real part of $\f{M}$ 
must have two positive eigenvalues $\lambda_{1,2}$. 

Now, the derivative of $\psi$ yields 
\bea
\frac{\partial}{\partial\f{x}}\,\psi(\f{x})=-\f{M}\cdot\f{x}\,\psi(\f{x})
\,.
\ea
In terms of the momentum operator $\hat{\f{p}}$, we get  
\bea
\label{appendix-annihilation}
\left(i\hat{\f{p}}+\f{M}\cdot\hat{\f{x}}
\right)\ket{\psi}=0
\,.
\ea
This motivates the introduction of pre-annihilation operators 
$\hat{\f{A}}=i\hat{\f{p}}+\f{M}\cdot\hat{\f{x}}$ which obey the 
following commutation relations
\bea
\left[\hat A_I,\hat A_J\right]
&=&
\left[\hat A_I^\dagger,\hat A_J^\dagger\right]
=0
\,,
\nn
\left[\hat A_I,\hat A_J^\dagger\right]
&=&
M_{IJ}+M_{IJ}^*=2\Re(M_{IJ})
\,.
\ea
Since $\Re(\f{M})$ is a real symmetric and positive matrix, we may
diagonalize it with an orthogonal (rotation) matrix $\f{D}$
such that
$\f{D}\cdot\Re(\f{M})\cdot\f{D}^\dagger={\rm diag}\{\lambda_I\}$.
As a result, the operators 
$\hat{\f{a}}=\f{D}\cdot[2\Re(\f{M})]^{-1/2}\cdot\hat{\f{A}}$,
i.e., 
\bea
\hat a_I=\frac{D_{IJ}\hat A_J}{\sqrt{2\lambda_I}}
\;\leadsto\;
\hat a_I\ket{\psi}=0
\ea
satisfy the standard commutation relations and do also annihilate the 
state $\ket{\psi}$.
Since this state $\ket{\psi}$ is just the initial vacuum state reduced to the 
two modes $\hat a_H$ and $\hat a_P$, the two operators
$\hat a_{1,2}$ above must be a linear combination of the initial 
annihilation operators.
>From the construction above, we see that the Hawking and partner mode 
operators $\hat a_H$ and $\hat a_P$ must be linear combinations of 
these operators $\hat a_1$ and $\hat a_2$ as well as their adjoints 
$\hat a_1^\dagger$ and $\hat a_2^\dagger$. 
Thus, the linear sub-space spanned by $\hat a_1$ and $\hat a_2$ 
can be identified with that of $\hat a_\|$ and $\hat a_\perp$ and 
we arrive at the ansatz~(\ref{partner-ansatz}).  

Alternatively, one could insert the general ansatz for the modes 
$\hat a_H
=
%\scalar{\f{\alpha}}{\hat{\f{a}}}+
%\left(\scalar{\f{\beta}}{\hat{\f{a}}}\right)^\dagger
%=
\scalar{\f{\alpha}}{\hat{\f{a}}}+\scalar{\hat{\f{a}}}{\f{\beta}}$
and 
$\hat a_P
=
%\scalar{\f{\alpha}}{\hat{\f{a}}}+
%\left(\scalar{\f{\beta}}{\hat{\f{a}}}\right)^\dagger
%=
\scalar{\f{\gamma}}{\hat{\f{a}}}+\scalar{\hat{\f{a}}}{\f{\delta}}$
into Eq.~(\ref{appendix-annihilation}) which gives the two linear equations
\bea
\f{\beta}-\f{\alpha}+M_{11}(\f{\beta}+\f{\alpha})+M_{12}(\f{\delta}+\f{\gamma}) &=& 0 
\,,
\nn
\f{\delta}-\f{\gamma}+M_{21}(\f{\delta}+\f{\gamma})+M_{22}(\f{\beta}+\f{\alpha}) &=& 0 
\,,
\ea
where $M_{IJ}$ are the components of the symmetric matrix $\f{M}$ 
(which also depend on $\f{\alpha}$, $\f{\beta}$,  $\f{\gamma}$, and  $\f{\delta}$). 
Since these two equations are linearly independent for all $M_{IJ}$, we find that 
$\f{\gamma}$ and  $\f{\delta}$ must lie in the same sub-space as 
$\f{\alpha}$ and $\f{\beta}$ (which are assumed to be linearly independent). 

\end{document}